\documentclass[acmsmall]{acmart}

\usepackage{multirow}
\usepackage{cellspace}
\usepackage{hyperref}
\usepackage{graphicx} 
\usepackage{tablefootnote}
\usepackage{hyperref}

\renewcommand{\thefootnote}{\fnsymbol{footnote}}
\newcommand{\numfootnote}[1]{\renewcommand{\thefootnote}{\arabic{footnote}}\footnote{#1}\renewcommand{\thefootnote}{\fnsymbol{footnote}}}

\AtBeginDocument{%
  }

\setcopyright{acmlicensed}
\copyrightyear{2025}
\acmYear{2025}
\acmDOI{XXXXXXX.XXXXXXX}

\acmISBN{978-1-4503-XXXX-X/18/06}

\begin{document}

\title{
Payload-Aware Intrusion Detection with CMAE and Large Language Models
}

\author{Yongcheol Kim}
\email{kimycn1017@gmail.com}
\orcid{0009-0000-6687-9845}
\affiliation{%
  \institution{Neouly Co., Ltd.}
  \city{Seoul}
  \country{South Korea}
}

\author{Chanjae Lee}
\email{arisel117@gmail.com}
\orcid{0000-0001-8838-6128}
\affiliation{%
  \institution{NetCoreTech Co., Ltd.}
  \city{Seoul}
  \country{South Korea}
}

\author{Young Yoon}
\email{young.yoon@hongik.ac.kr}
\orcid{0000-0002-5249-2823}
\authornote{Corresponding Author}
\affiliation{%
  \institution{Hongik University}
  \city{Seoul}
  \country{South Korea}
}

\renewcommand{\shortauthors}{Kim, Lee, Yoon}

\begin{abstract}

Intrusion Detection Systems (IDS) are crucial for identifying malicious traffic, yet traditional signature-based methods struggle with zero-day attacks and high false positive rates. AI-driven packet-capture analysis offers a promising alternative. However, existing approaches rely heavily on flow-based or statistical features, limiting their ability to detect fine-grained attack patterns.

This study proposes Xavier-CMAE, an enhanced Convolutional Multi-Head Attention Ensemble (CMAE) model that improves detection accuracy while reducing computational overhead. By replacing Word2Vec embeddings with a Hex2Int tokenizer and Xavier initialization, Xavier-CMAE eliminates pre-training, accelerates training, and achieves 99.971\% accuracy with a 0.018\% false positive rate, outperforming Word2Vec-based methods.

Additionally, we introduce LLM-CMAE, which integrates pre-trained Large Language Model (LLM) tokenizers into CMAE. While LLMs enhance feature extraction, their computational cost hinders real-time detection. LLM-CMAE balances efficiency and performance, reaching 99.969\% accuracy with a 0.019\% false positive rate.

This work advances AI-powered IDS by (1) introducing a payload-based detection framework, (2) enhancing efficiency with Xavier-CMAE, and (3) integrating LLM tokenizers for improved real-time detection.

\end{abstract}



\received{}
\received[revised]{}
\received[accepted]{}

\maketitle

\section{Introduction}\label{sec:intro}

\subsection{The Necessity and Limitations of AI in Network Security}\label{sec:ai_ids}

Intrusion Detection Systems (IDS) play a crucial role in identifying malicious traffic within network environments~\cite{rains2020cybersecurity, wangen2016cyber, akshaya2019study, patil2024leveraging}. The widely used signature-based detection approach ensures high detection performance for known attack patterns; however, it struggles to detect emerging threats such as zero-day attacks~\cite{liao2013intrusion, deldar2023deep}. In addition, maintaining complex rule sets requires significant effort, and the system is prone to false positives. To overcome these limitations, machine learning and deep learning-based detection techniques have been actively explored~\cite{ahmed2024delm}, particularly focusing on analyzing packet captures to enhance malicious traffic detection.

AI-driven approaches have recently emerged as powerful tools to address these challenges~\cite{li2018cyber, sarker2021ai, wirkuttis2017artificial, morel2011artificial, das2021artificial, ansari2022impact}. By leveraging AI, automated feature extraction and pattern recognition enable the reduction of false positives and improve the detection of new attacks~\cite{vigneswaran2018evaluating, podder2021artificial, navya2021intrusion, liu2020deep, akgun2022new, kim2020cnn, alabadi2020anomaly, ding2018intrusion, kim2019intrusion, cao2022network, al2023intrusion, koniki2022anomaly, ansari2022gru, wu2022rtids, ullah2024ids, MallampatiSeetha+2024+98+117, kim2023ensemble}.

However, most existing studies primarily rely on flow-based statistical features for intrusion detection. While this approach may achieve high detection rates for specific attack types, it lacks the granularity required for precise threat identification since it does not directly analyze payload contents. The Transformer~\cite{vaswani2017attention}-based Convolutional Multi-Head Attention Ensemble (CMAE) model~\cite{kim2023ensemble} has demonstrated strong performance by directly learning and analyzing packet payload data to improve detection accuracy.

Despite these advantages, even a single missed detection in cybersecurity can result in severe consequences, highlighting the need for further improvements in the CMAE model~\cite{kim2023ensemble}. A major limitation of CMAE~\cite{kim2023ensemble} is its dependence on pre-trained Word2Vec embeddings, which not only increases training time but also poses difficulties in finding optimal embedding distributions, particularly as payload length increases. To address these challenges, this study discusses the strengths and weaknesses of CMAE~\cite{kim2023ensemble} and proposes an enhanced approach to improve its performance.

\subsection{Research Trends on IDS Utilizing Large Language Models (LLMs)}\label{sec:llm_trend}

The adoption of large language models (LLMs)~\cite{radford2018improving, touvron2023Llama1, touvron2023Llama2, dubey2024Llama, le2023bloom, almazrouei2023falcon, jiang2024mixtral, basyal2023text} in various domains has gained significant attention due to their ability to learn from vast datasets and recognize complex patterns. Attempts have been made to integrate LLMs~\cite{touvron2023Llama2, dubey2024Llama} into IDS; however, several challenges remain unresolved.

First, open LLMs~\cite{touvron2023Llama2, dubey2024Llama} that are pre-trained on general domain data could be exploited to generate cyberattacks if fine-tuned on security-related datasets, raising concerns about their direct application in cybersecurity~\cite{yao2024survey}.

\begin{table}[!htbp]
    \centering
    \caption{Unencrypted Payload Classification Results Using Various LLMs on the CIC-IDS2017 Dataset}
    
    \begingroup
    \setlength{\tabcolsep}{3pt} 
    \renewcommand{\arraystretch}{1.1} 
    \resizebox{\textwidth}{!}{ 

        \begin{tabular}{c|ccccccc|c}
        \toprule[1.5pt]

        \textbf{Model} &
            \textbf{Benign} & \textbf{Dos} & \textbf{DDoS} &
            \textbf{\begin{tabular}[c]{@{}c@{}}Port\\[-2pt] Scan\end{tabular}} &
            \textbf{\begin{tabular}[c]{@{}c@{}}Brute\\[-2pt] Force\end{tabular}} &
            \textbf{Bot} & \textbf{Web} &
            \textbf{\begin{tabular}[c]{@{}c@{}}Number of\\[-2pt] Correct\end{tabular}} \\
        \hline \hline

        GPT-4o~\cite{hurst2024gpt} & Web & Web & Web & - & Brute Force & Benign & Web & 2 \\
        GPT-4o mini~\cite{hurst2024gpt} & Web & Web & Benign & - & Benign & Web & Web & 1 \\

        Llama2-7B~\cite{touvron2023Llama2} & Dos & Dos & Dos & - & Dos & Dos & Dos & 1 \\
        Llama3-7B~\cite{dubey2024Llama} & Web & Web & Web & - & Web & Web & Web & 1 \\

        Claude 3.7 Sonnet~\numfootnote{\href{https://claude.ai/}{https://claude.ai/}} & Benign & Bot & Port Scan & - & Brute Force & Web & Bot & 2 \\

        Grok3~\numfootnote{\href{https://grok.com/}{https://grok.com/}} & Web & Web & Web & - & Bot & Web & Web & 1 \\

        \bottomrule[1.5pt]
        \end{tabular}
    }
    \endgroup
    \label{tab:payload_qeustion}
\end{table}

\autoref{tab:payload_qeustion} presents the results of querying various LLMs with randomly extracted unencrypted payloads from the CIC-IDS2017 dataset. The question format used to generate responses is as follows:

\textbf{"\_\_Question\_Payload\_\_" This string is a payload captured from an IDS device. Please select only one of ["Benign", "DoS", "DDoS", "Port Scan", "Brute Force", "Bot", "Web"] and answer without any other questions.}

\begin{enumerate}
    \item \textbf{Benign}
    \begin{itemize}
        \item "474554202f4d464577547a424e4d4573775354414a4267557244674d4347675541424253353662 4b48416f55442532424f796c253242304c685067394a7879516d3467515566394e6c70384c64374c 76774d416e7a517a6e364171387a4d544d4345483768536d3976372532464c54667a253242745a5 5303632725125334420485454502f312e310d0a436f6e6e656374696f6e3a204b6565702d416c697 6650d0a4163636570743a202a2f2a0d0a557365722d4167656e743a204d6963726f736f66742d437 27970746f4150492f31302e300d0a486f73743a2073322e73796d63622e636f6d0d0a0d0a"
    \end{itemize}

    \item \textbf{DoS}
    \begin{itemize}
        \item "474554202f20485454502f312e310d0a486f73743a203230352e3137342e3136352e36380d0a557 365722d4167656e743a204d6f7a696c6c612f342e302028636f6d70617469626c653b204d534945 20372e303b2057696e646f7773204e5420352e313b2054726964656e742f342e303b202e4e45542 0434c5220312e312e343332323b202e4e455420434c5220322e302e3530336c333b202e4e455420 434c5220332e302e343530362e323135323b202e4e455420434c5220332e352e33303732393b204 d534f6666696365203132290d0a436f6e74656e742d4c656e6774683a2034320d0a"
    \end{itemize}

    \item \textbf{DDoS}
    \begin{itemize}
        \item "474554202f20485454502f312e300d0a0d0a0d0a"
    \end{itemize}

    \item \textbf{Brute Force}
    \begin{itemize}
        \item "5353482d322e302d706172616d696b6f5f322e302e300d0a"
    \end{itemize}

    \item \textbf{Web}
    \begin{itemize}
        \item "474554202f64762f20485454502f312e310d0a486f73743a203230352e3137342e3136352e36380 d0a557365722d4167656e743a204d6f7a696c6c612f352e3020285831313b204c696e75782078383 65f36343b2072763a34352e3029204765636b6f2f32303130303130312046697265666f782f34352e 300d0a4163636570743a20746578742f68746d6c2c6170706c69636174696f6e2f7868746d6c2b78 6d6c2c6170706c69636174696f6e2f786d6c3b713d302e392c2a2f2a3b713d302e380d0a41636365 70742d4c616e67756167653a20656e2d55532c656e3b713d302e350d0a4163636570742d456e636 f64696e673a20677a69702c206465666c6174650d0a436f6e6e656374696f6e3a206b6565702d616c 6976650d0a0d0a"
    \end{itemize}

    \item \textbf{Bot}
    \begin{itemize}
        \item "485454502f312e3120323030204f4b0d0a446174653a204672692c203037204a756c20323031372 031333a31303a333220474d540d0a436f6e74656e742d4c656e6774683a20300d0a436f6e74656e 742d547970653a20746578742f68746d6c3b636861727365743d7574662d380d0a5365727665723 a20417265730d0a0d0a"
    \end{itemize}

\end{enumerate}

In this experiment, port-scan attacks were excluded because all corresponding payloads were encrypted. The evaluation, conducted on March 7, 2025, showed that the best-performing models, GPT-4o and Claude 3.7 Sonnet, correctly classified only 2 out of 6 classes. Overall, LLMs exhibited low classification accuracy, ranging from 16\% to 33\%. The results indicate that without access to external knowledge, LLMs frequently misclassify payloads, often assigning them to incorrect categories.

Accurately identifying cyberattacks requires deep domain knowledge in cybersecurity, such as traffic patterns and attack sequences. However, LLMs lack dedicated training on security-specific data or access to such knowledge, which likely contributed to their poor performance. Unlike traditional network intrusion detection methods, which leverage behavioral patterns and anomaly detection, LLMs primarily rely on natural language understanding. As a result, they tend to misinterpret structured or encoded payloads, leading to frequent misclassification of attack types and benign traffic.

Furthermore, the absence of reasoning-based techniques, such as Reasoning-oriented RL~\cite{guo2025deepseek} and Reasoning Augmented Generation (ReAG)~\numfootnote{\href{https://github.com/superagent-ai/reag}{https://github.com/superagent-ai/reag}}, may also have contributed to the performance gap. These approaches allow models to systematically break down complex problems and enhance contextual reasoning. However, the LLMs tested in this experiment did not employ such methods and instead relied on superficial pattern matching. Consequently, their classification decisions were often inaccurate, highlighting the need for domain-adapted LLM training strategies and enhanced reasoning capabilities to improve cybersecurity applications.

Second, applying LLMs~\cite{touvron2023Llama2, dubey2024Llama} to IDS requires fine-tuning on domain-specific security data, which involves significant costs in terms of data collection, preprocessing, environment setup, human-in-the-loop annotation, model training, and tuning~\cite{kheddar2024transformers}. Due to these constraints, LLM-based IDS remains an evolving research area. This study aims to explore the strengths and limitations of LLMs~\cite{touvron2023Llama2, dubey2024Llama} in this context and propose feasible solutions to enhance their applicability.

\subsection{Proposed Approach and Key Contributions}\label{sec:proposed}

To address the limitations of the existing CMAE model~\cite{kim2023ensemble}, this study proposes the Xavier-initialized CMAE (Xavier-CMAE) model. Xavier-CMAE employs the Hex2Int tokenizer to transform packet payload data while initializing the embedding layer using Xavier initialization. This design eliminates the need for separate Word2Vec~\cite{mikolov2013efficient} training, allowing for more effective feature learning. By doing so, the model significantly enhances training speed while ensuring that the embedding vectors maintain an appropriate distribution as payload length increases. The experimental results demonstrate that the proposed method achieves a detection accuracy of 99.9718\% and a false positive rate of 0.0182\%, outperforming conventional Word2Vec~\cite{mikolov2013efficient}-based methods.

Additionally, while lightweight adaptation techniques such as LoRA~\cite{hu2022lora} have been introduced to mitigate the high computational and training costs of LLMs~\cite{touvron2023Llama2, dubey2024Llama}, this study proposes LLM-CMAE, an LLM-enhanced CMAE model, to address these challenges even further. LLM-CMAE integrates pre-trained LLM tokenizers and embeddings into CMAE~\cite{kim2023ensemble}, leveraging the expressive power of LLMs while maintaining CMAE’s real-time efficiency. This integration enhances feature extraction and detection performance, achieving 99.97\% accuracy with a 0.019\% false positive rate, while maintaining efficient training and inference speeds.

The key contributions of this study are as follows:

\begin{itemize}
    \item Payload-based AI Detection Approach: Unlike conventional IDS approaches that primarily rely on flow-based statistical features from packet captures, this study applies AI-driven techniques to directly analyze payload data for malicious packet detection.
    \item Xavier-CMAE Model: This study proposes an improved CMAE model that replaces Word2Vec embeddings with Hex2Int tokenization and Xavier initialization, enhancing both training efficiency and detection performance.
    \item LLM-CMAE Model: To overcome the computational overhead and domain knowledge limitations of pre-trained LLMs~\cite{touvron2023Llama2, dubey2024Llama}, this study integrates LLM tokenizers and embedding layers into CMAE~\cite{kim2023ensemble}, ensuring high detection accuracy while maintaining real-time processing capabilities.
\end{itemize}

While previous studies have focused on statistical feature-based intrusion detection, this study introduces an AI-based detection framework centered on the analysis of payload content. Through this approach, we demonstrate that more precise detection of malicious network traffic can be achieved.

\section{Related Works}\label{sec:related}

\subsection{Advances in AI-Based Intrusion Detection Systems}

Intrusion Detection Systems (IDS) have become a crucial technology in network security, enabling the identification and mitigation of malicious traffic~\cite{rains2020cybersecurity, wangen2016cyber, akshaya2019study, patil2024leveraging}. Traditional IDS primarily utilize signature-based detection mechanisms to recognize known attack patterns, with Snort being a representative example~\cite{caswell2003snort, soe2019rule, kwon2022advanced, einy2021anomaly}. While this approach ensures high detection accuracy, it struggles to identify unknown (zero-day) attacks and often results in a high false positive rate, increasing the burden on security operators~\cite{liao2013intrusion, deldar2023deep}.

To overcome these limitations, various studies have explored the integration of artificial intelligence (AI) into cybersecurity, as seen in Network Detection and Response (NDR) solutions, which highlight the necessity and direction of AI applications in this domain~\cite{li2018cyber, sarker2021ai, wirkuttis2017artificial, morel2011artificial, das2021artificial, ansari2022impact}. Among traditional machine learning approaches applied to cybersecurity, methods such as Random Forest~\cite{choubisa2022simple, kilincer2021machine}, SVM~\cite{kilincer2021machine, hong2021machine, avci2023cybersecurity}, kNN~\cite{kilincer2021machine, hong2021machine, avci2023cybersecurity}, and XGBoost~\cite{gouveia2020network, dhaliwal2018effective, leevy2021detecting} have been widely used to detect threats by learning specific attack patterns.

Further advancements in deep learning have led to the development of models that automatically learn complex attack patterns from network traffic data. Deep Neural Networks (DNNs)\cite{hinton2006fast} have been employed in various IDS frameworks\cite{vigneswaran2018evaluating, podder2021artificial, navya2021intrusion, liu2020deep, akgun2022new}, leveraging statistical packet data to enhance detection performance. Convolutional Neural Networks (CNNs)\cite{lecun1989backpropagation} have also been explored for intrusion detection\cite{kim2020cnn, alabadi2020anomaly, ding2018intrusion, kim2019intrusion}, as they effectively capture local features in network traffic. However, CNNs are limited in capturing long-term dependencies, prompting the adoption of Recurrent Neural Networks (RNNs), including Gated Recurrent Units (GRUs)\cite{cho2014learning} and Long Short-Term Memory (LSTM) networks\cite{hochreiter1997long}~\cite{cao2022network, al2023intrusion, koniki2022anomaly, ansari2022gru}. Despite their effectiveness, these recurrent models require substantial computational resources and suffer from vanishing gradient issues, making them difficult to train for long sequences.

More recently, Transformer-based architectures~\cite{vaswani2017attention} have been introduced to address long-term dependency issues, demonstrating superior performance in IDS applications~\cite{wu2022rtids, ullah2024ids, MallampatiSeetha+2024+98+117}. However, the majority of existing AI-based IDS approaches primarily rely on statistical features extracted from packet headers, network flows, and metadata. These feature-engineering processes are often labor-intensive and limit generalization to novel threats. Furthermore, by not directly analyzing raw payloads, these methods remain vulnerable to adversarial attacks that embed malicious code within payload data, which statistical analysis alone may fail to detect~\cite{chakkaravarthy2019survey, vidal2020espada}.

A promising alternative is the Convolutional Multi-Head Attention Ensemble (CMAE) model, which has recently gained attention as an efficient deep learning framework for IDS~\cite{kim2023ensemble}. By combining CNNs and Transformer-based attention mechanisms, CMAE achieves high detection performance while reducing computational overhead. Moreover, its ability to directly process payload data allows it to surpass traditional statistical approaches. However, CMAE still has room for improvement, particularly in real-time intrusion detection scenarios that demand both high accuracy and computational efficiency.

\subsection{Large Language Models (LLMs) and Their Potential in Network Security}

Large Language Models (LLMs) have made significant advancements in natural language processing (NLP) and are increasingly being applied to various domains~\cite{vakayil2024rag, zhang2023sentiment, saharia2022photorealistic}. In cybersecurity, LLMs offer new possibilities for automating threat intelligence and response systems. Their ability to process and generate human-like text allows them to extract rich contextual features, which can be valuable for network traffic analysis. By leveraging pre-trained tokenizers and embeddings, LLMs can learn high-dimensional feature representations, potentially enhancing IDS capabilities.

Despite their strengths, integrating LLMs into IDS frameworks presents several challenges. First, LLMs contain billions of parameters, making real-time intrusion detection computationally expensive~\cite{kheddar2024transformers}. Second, applying LLMs to IDS requires domain-specific training, which raises ethical and data privacy concerns, particularly when handling sensitive security data~\cite{yao2024survey}. Moreover, while LLMs can learn complex security concepts, they also risk unintentionally encoding adversarial knowledge, potentially generating harmful content such as cyberattack strategies. Due to these ethical constraints, LLMs are often trained with restricted datasets, limiting their effectiveness in security applications.

To address these issues, several model optimization techniques have been explored, including knowledge distillation\cite{xu2024survey}, quantization\cite{lin2024awq}, model compression\cite{kwon2022fast}, and low-rank factorization\cite{saha2025compressing}. Alternatively, LLMs can be adapted as feature extractors, utilizing pre-trained backbone models such as ResNet~\cite{he2016deep}, Vision Transformer (ViT)\cite{dosovitskiy2020image}, BERT\cite{devlin2019bert}, T5~\cite{raffel2020exploring}, GPT~\cite{radford2018improving}, and LLaMA~\cite{touvron2023Llama1, touvron2023Llama2, dubey2024Llama}. Several studies have demonstrated the effectiveness of such an approach across different domains~\cite{dai2021up, dosovitskiy2020image, liu2019roberta}. This suggests that LLM-based tokenization and embedding strategies can be leveraged for IDS, enabling efficient and expressive feature extraction while maintaining computational efficiency.

\subsection{Comparison with Existing Studies and Contributions of This Research}

\begin{table}[!htbp]
    \centering
    \caption{Comparison of Detection Methods}
    \begingroup
    \setlength{\tabcolsep}{3pt} 
    \renewcommand{\arraystretch}{1.1} 
    \resizebox{\textwidth}{!}{ 
        \begin{tabular}{c|ccc}
        \toprule[1.5pt]

        \textbf{Method} & \textbf{Characteristics} & \textbf{Advantages} & \textbf{Disadvantages} \\ \hline \hline

        \textbf{Signature Snort~\cite{caswell2003snort}} & \begin{tabular}[c]{@{}c@{}}Static rule-based\\ packet detection\end{tabular} & \begin{tabular}[c]{@{}c@{}}Low computational cost\\ Real-time detection\end{tabular} & \begin{tabular}[c]{@{}c@{}}High false positive rate\\ Unable to detect zero-day attacks\end{tabular} \\ \hline

        \begin{tabular}[c]{@{}c@{}}\textbf{Deep Learning}\\ ~\cite{vigneswaran2018evaluating, podder2021artificial, navya2021intrusion, liu2020deep, akgun2022new, kim2020cnn, alabadi2020anomaly, ding2018intrusion} \\ ~\cite{kim2019intrusion, cao2022network, al2023intrusion, koniki2022anomaly, ansari2022gru, wu2022rtids, ullah2024ids, MallampatiSeetha+2024+98+117} \end{tabular} &

        \begin{tabular}[c]{@{}c@{}}Learning\\ packet statistics\\ and patterns\end{tabular} & \begin{tabular}[c]{@{}c@{}}Automated detection\\ High detection accuracy\end{tabular} & \begin{tabular}[c]{@{}c@{}}Requires feature engineering\\ Real-time inference overhead\\ Limited malware detection\end{tabular} \\ \hline

        \textbf{CMAE~\cite{kim2023ensemble}} &
        \begin{tabular}[c]{@{}c@{}}Payload-based\\ detection\end{tabular} & \begin{tabular}[c]{@{}c@{}}High detection performance\\ Automated real-time detection\\ Effective malware detection\end{tabular} & Potential for further optimization \\ \hline

        \textbf{LLM~\cite{touvron2023Llama2, dubey2024Llama}} &
        \begin{tabular}[c]{@{}c@{}}Adaptive detection\\ via fine-tuning\end{tabular} & \begin{tabular}[c]{@{}c@{}}Strong representation learning\\ Domain adaptation\\ Effective malware detection\end{tabular} & \begin{tabular}[c]{@{}c@{}}High computational cost\\ Difficulties in real-time detection\\ Limited training on sensitive data\end{tabular} \\ \hline

        \textbf{\begin{tabular}[c]{@{}c@{}}LLM-CMAE\\ (Ours)\end{tabular}} &
        \begin{tabular}[c]{@{}c@{}}Integrating\\ pre-trained LLM\\ into CMAE\end{tabular} & \begin{tabular}[c]{@{}c@{}}High detection performance\\ Automated detection\\ Effective malware detection\end{tabular} & \begin{tabular}[c]{@{}c@{}}Requires real-time optimization\\ Limited training on sensitive data\end{tabular} \\ \hline

        \textbf{\begin{tabular}[c]{@{}c@{}}Xavier-CMAE\\ (Ours)\end{tabular}} &
        \begin{tabular}[c]{@{}c@{}}Optimized CMAE\\ for improved efficiency\end{tabular} & \begin{tabular}[c]{@{}c@{}}High detection performance\\ Automated real-time detection\\ Effective malware detection\\ Optimized computational efficiency\end{tabular} & \begin{tabular}[c]{@{}c@{}}Potential degradation\\ in representation learning\\ performance\end{tabular} \\

        \bottomrule[1.5pt]
        \end{tabular}
    }
    \endgroup
    \label{tab:comparison}
\end{table}

To highlight the distinctions between existing approaches and our research, we summarize key intrusion detection techniques in \autoref{tab:comparison}. Traditional signature-based detection provides low computational cost and fast detection speed but suffers from a high false positive rate and an inability to detect zero-day attacks. Deep learning-based methods improve detection accuracy by learning statistical patterns in network traffic; however, they require extensive feature engineering and impose significant inference costs. CMAE~\cite{kim2023ensemble}-based approaches leverage payload data for fine-grained attack detection, proving effective in identifying malware embedded within payloads. While LLM~\cite{touvron2023Llama2, dubey2024Llama}-based detection methods exhibit strong feature extraction capabilities and transferability across different environments, their high computational requirements pose challenges for real-time IDS deployment.

To address these limitations, our research enhances CMAE~\cite{kim2023ensemble} by refining its tokenizer and embedding layer to strike a balance between detection performance and computational efficiency. Specifically, we introduce Xavier-CMAE, which replaces Word2Vec~\cite{mikolov2013efficient} embeddings with a Hex2Int tokenizer and Xavier initialization, eliminating pre-training overhead while ensuring well-structured vector representations. Additionally, we propose LLM-CMAE, integrating pre-trained LLM~\cite{touvron2023Llama2, dubey2024Llama} tokenizers into CMAE~\cite{kim2023ensemble} to leverage LLMs~\cite{touvron2023Llama2, dubey2024Llama}’ expressiveness while maintaining computational efficiency. By doing so, we aim to enhance both accuracy and real-time feasibility for IDS applications.

\section{Methodology}\label{sec:method}

\subsection{Overview of the CMAE Model}\label{sec:cmae}

\begin{figure}[!htbp]
  \centering
  \includegraphics[width=\linewidth]{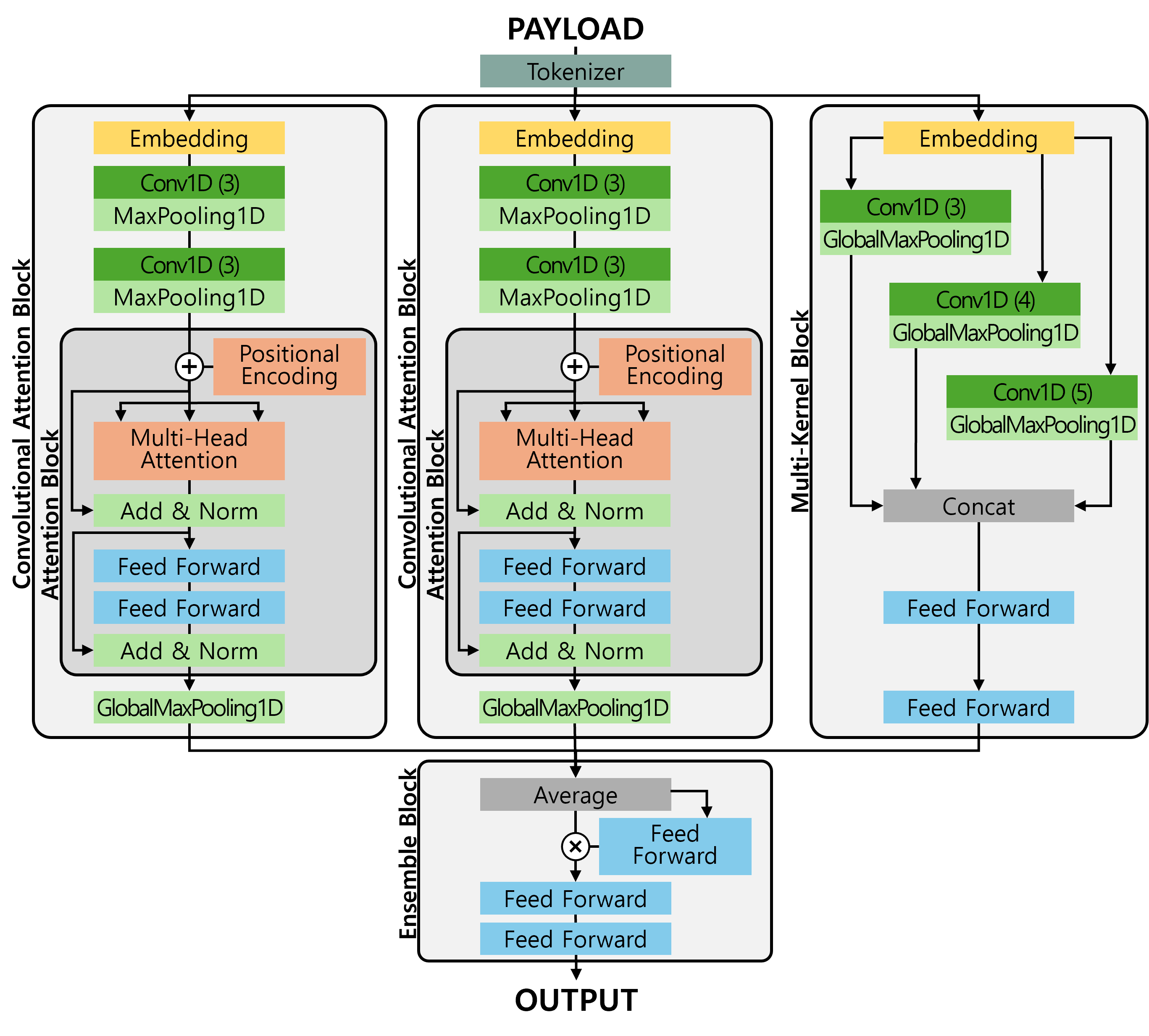}
  \caption{Bseline Model: CMAE Architecture}
  \Description{CMAE Model Architecture}
  \label{fig:came_architecture}
\end{figure}

In this study, we utilize the Convolutional Multi-Head Attention Ensemble (CMAE) model~\cite{kim2023ensemble} as the baseline, with its overall architecture illustrated in \autoref{fig:came_architecture}. The model is designed to tokenize payload data using a tokenizer and input each token individually into multiple processing blocks. The CMAE model~\cite{kim2023ensemble} comprises two Convolutional Attention Blocks (CABs) and one Multi-Kernel Block (MKB). The outputs from each block are integrated via an Average Layer and used for final predictions.

\subsubsection{\textbf{Convolutional Attention Block}}\label{sec:cmae_cab}

\leavevmode\newline
The CAB structure sequentially applies a Conv1D layer with a kernel size of 3, followed by a MaxPooling1D layer twice. Subsequently, the data passes through an Attention Block and is finally processed by a GlobalMaxPooling1D layer to generate the final output.

The Attention Block is based on the Encoder Block structure~\cite{vaswani2017attention}. It first adds a positional encoding vector to the input tensor, which then passes through a Multi-Head Attention Layer. A skip connection~\cite{he2016deep} is employed to merge the original input with the attention output, followed by normalization. The output is then processed through two Feed-Forward Layers, after which another skip connection and Layer Normalization~\cite{ba2016layer} are applied. Finally, the output is reduced to a lower-dimensional representation using a GlobalMaxPooling1D Layer.

Prior to being fed into the Attention Block, the CAB utilizes a 1D Convolutional Layer and a MaxPooling Layer to project input data into a lower-dimensional space. This process reduces memory usage and enhances computational speed. Additionally, the 1D Convolutional Layer ensures that individual characters are recognized as meaningful tokens, aligning with the characteristics of an IDS environment. The lower-dimensional tensor is then processed by the Self-Attention Layer~\cite{vaswani2017attention}, maintaining relationships between distant inputs.

\subsubsection{\textbf{Multi-Kernel Block}}\label{sec:cmae_mkb}

\leavevmode\newline
The MKB applies Conv1D layers (1D Convolutional Layer) with different kernel sizes (3, 4, and 5) in parallel. Each output is then reduced using a GlobalMaxPooling1D Layer, after which the results are concatenated into a single tensor. This concatenated tensor is subsequently processed through two consecutive Feed-Forward Layers to generate the final output.

MKB incorporates the Multi-Scale Convolution Layer concept from the Inception Model~\cite{szegedy2015going}, enabling the extraction of features at various levels by applying different kernel sizes in parallel. This design reduces computational complexity while preventing excessive reliance on CAB, thereby improving the model’s generalization performance.

\subsubsection{\textbf{Ensemble Block}}\label{sec:cmae_eb}

\leavevmode\newline
The Ensemble Block effectively integrates features extracted from preceding blocks using an ensemble technique. The three outputs obtained from the two CABs and one MKB are first combined using an Average Layer.

The aggregated output undergoes a non-linear transformation through a Feed-Forward Layer. A gating mechanism similar to the Attention Mechanism~\cite{vaswani2017attention} is then applied, performing element-wise scaling between the original input and the processed output to emphasize significant features. This mechanism enables selective activation of relevant features. Finally, the transformed tensor passes through a Feed-Forward Layer, where the output dimension is adjusted for classification. The final probability distribution is generated using the Softmax function~\cite{nwankpa2018activation}.

\subsubsection{\textbf{Performance and Limitations}}\label{sec:cmae_limit}

\leavevmode\newline
The CMAE model~\cite{kim2023ensemble} achieved an accuracy of 99.82 \% in multi-class classification experiments using the CIC-IDS2017 dataset, with an inference speed exceeding 5,000 packets per second. These results demonstrate its effectiveness in IDS environments.

However, continuous optimization is required to enhance detection performance in IDS applications. In environments where over one million malicious packets are analyzed daily, even a 1\% improvement in accuracy can significantly reduce the number of false positives (FP) and false negatives (FN) by millions annually. Furthermore, IDS systems can optimize operational strategies in the following ways:

\begin{itemize}
    \item \textbf{High-Accuracy, Low-Speed Models:} These models can be used to identify high-risk threats or detect zero-day attacks, where false negatives are particularly critical.
    \item \textbf{Low-Accuracy, High-Speed Models:} These models can be deployed in parallel before primary detection stages to pre-filter large volumes of data, thereby reducing false positives.
\end{itemize}

To address these challenges, this study proposes two novel approaches to enhance the performance of the CMAE model.

\subsection{LLM as a Backbone for CMAE}\label{sec:llm_cmae}

In an Intrusion Detection System (IDS) environment, high-performance real-time detection is essential, necessitating both high accuracy and fast inference speed. However, directly applying Large Language Models (LLMs) presents two major challenges:

\begin{itemize}
    \item \textbf{Decreased Computational Speed:} Due to their large tokenizers, embedding layers, and model size, LLMs exhibit relatively slow computational speed, making it difficult to meet the real-time processing performance required in IDS environments.
    \item \textbf{Limitations of Domain-Specific Data:} Network security-related data are often excluded from general LLM training processes due to concerns about potential misuse and ethical issues. Consequently, LLMs inherently lack the ability to learn the critical information necessary for IDS detection.
\end{itemize}

To address these challenges, this study proposes an approach that integrates the tokenizer and embedding layers—key components of an LLM—with the CMAE model. This methodology aligns with existing approaches that utilize backbone models such as ResNet~\cite{he2016deep}, ViT~\cite{dosovitskiy2020image}, and BERT~\cite{devlin2019bert} as feature extractors~\cite{dai2021up, dosovitskiy2020image, liu2019roberta}. Our approach optimizes this integration specifically for IDS applications.

The proposed CMAE model, which employs an LLM as its backbone~\cite{kim2023ensemble}, is designed to accommodate the constrained nature of the IDS domain. Its fundamental structure remains similar to the CMAE model introduced in \autoref{fig:came_architecture}\cite{kim2023ensemble}. The primary distinction lies in the modified tokenizer and embedding layers of the LLM, as depicted in \autoref{fig:subword_tokenizer} and \autoref{fig:hex_tokenizer}, while all other configurations adhere to the default CMAE model settings\cite{kim2023ensemble}. The integration of the LLM’s tokenizer and embedding layers with the CMAE model can be categorized into two approaches based on the input processing method.

The first approach, illustrated in \autoref{fig:subword_tokenizer}, utilizes the default input-output mechanism of LLMs~\cite{touvron2023Llama2, dubey2024Llama}. This method retains the original tokenizer and embedding layer of the pre-trained LLM model~\cite{touvron2023Llama2, dubey2024Llama} to maximize its rich representational capacity. By applying the LLM’s inherent representation learning capability to the IDS domain, this approach aims to maintain the efficiency of the existing CMAE model while enhancing accuracy and pattern recognition for IDS applications.

The second approach, shown in \autoref{fig:hex_tokenizer}, adapts the LLM’s tokenizer and embedding layers to align with the default input-output structure of the CMAE model~\cite{kim2023ensemble}. This method preserves the original input and output format of the CMAE model while integrating the LLM’s advanced representation learning capabilities to optimize performance for the IDS domain. By efficiently incorporating the strengths of LLMs, this approach seeks to maintain computational efficiency while ensuring domain-specific adaptability and improved detection performance.

\subsubsection{\textbf{Strategy 1: Sub-Word Tokenization and Embedding}}\label{sec:subword_token}

\begin{figure}[!htbp]
  \centering
  \includegraphics[height=7cm, keepaspectratio]{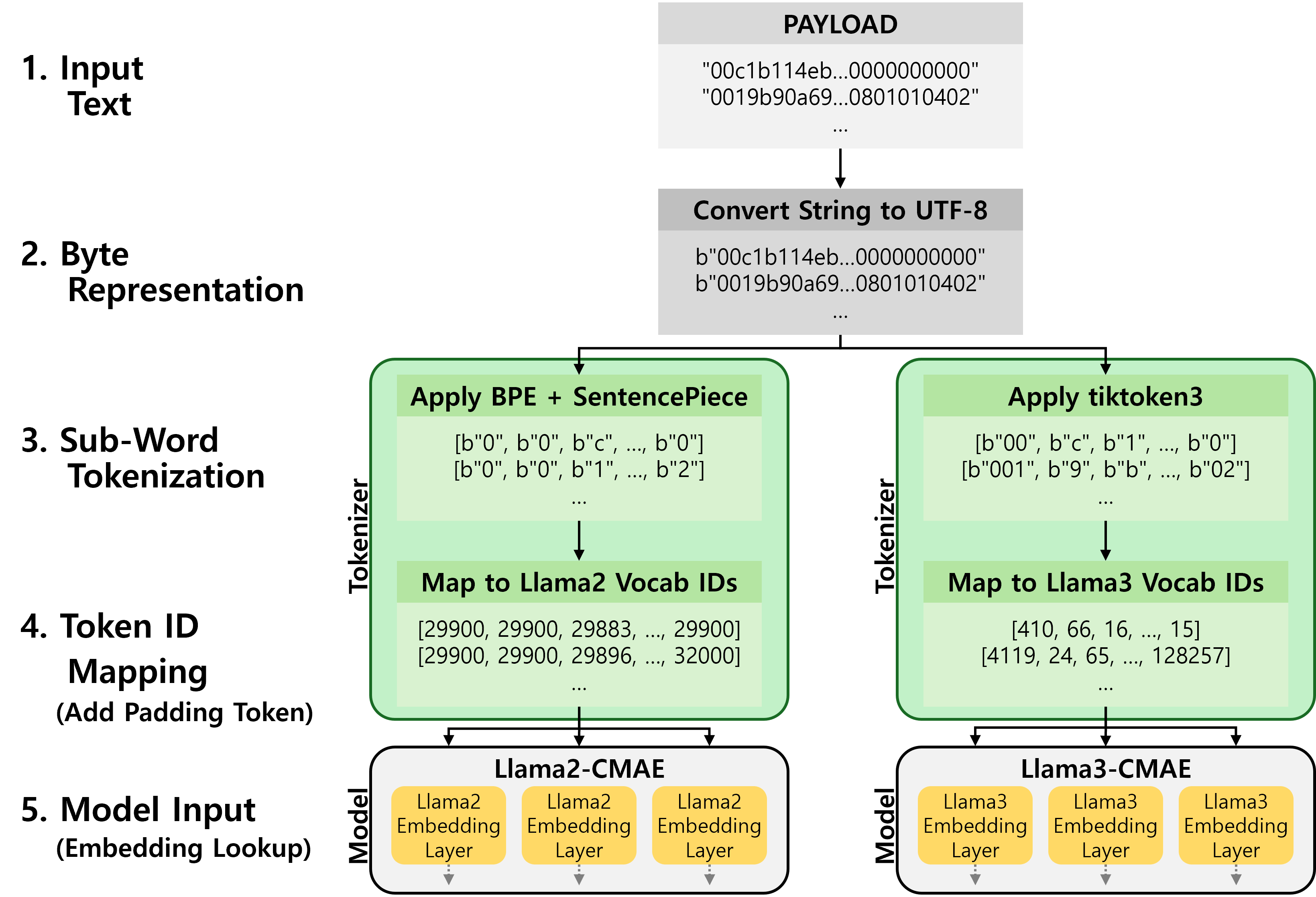}
  \caption{Sub-Word Tokenization}
  \Description{Sub-Word Tokenization}
  \label{fig:subword_tokenizer}
\end{figure}

\leavevmode\newline
The first approach involves tokenizing the payload data of network packets at the sub-word level using the Llama2-7B model~\cite{touvron2023Llama2} and the Llama3-8B model~\cite{dubey2024Llama}. As illustrated in steps 1 and 2 of \autoref{fig:subword_tokenizer}, this approach first converts the input string-type payload data into UTF-8 format and then tokenizes it into sub-word units for model input. In this method, the embedding layer is frozen to preserve the core representational learning capability of the model.

First, considering the common transformation steps shown in \autoref{fig:subword_tokenizer}, when a string such as "00c1b114eb…0000000000" is input in step 1, it is converted into UTF-8 format as b"00c1b114eb…0000000000" in step 2. The data is then processed according to each approach.

\begin{itemize}
    \item \textbf{Llama2-CMAE Model Based on BPE+SP Tokenizer}
    \begin{itemize}
        \item As shown on the left side of \autoref{fig:subword_tokenizer}, the payload converted into UTF-8 format is tokenized into sub-word units using the Byte Pair Encoding (BPE) algorithm~\cite{sennrich2015neural} and SentencePiece~\cite{kudo2018sentencepiece}. Then, the corresponding token IDs are retrieved from the vocabulary and mapped accordingly. Padding tokens with zero weights are added to match the input structure of the CMAE model~\cite{kim2023ensemble}. Furthermore, the three embedding layers on the front end of the CMAE model adopt the embedding layer of Llama2-7B~\cite{touvron2023Llama2}.
    
        \item This process is illustrated in \autoref{fig:subword_tokenizer} as follows. Applying BPE~\cite{sennrich2015neural} and SentencePiece~\cite{kudo2018sentencepiece} sequentially to the UTF-8 converted data results in a list of sub-words such as [b"0", b"0", b"c",..., b"0"] in Step 3 (left). Next, the token IDs corresponding to each word are retrieved from the vocabulary, yielding a mapped sequence like [29900, 29900, 29883, …, 29900] in Step 4. Finally, this sequence is structured as the model input in Step 5.
        
    \end{itemize}
    
    \item \textbf{Llama3-CMAE Model Based on tiktoken3 Tokenizer}
    \begin{itemize}
        \item As shown on the right side of \autoref{fig:subword_tokenizer}, this method utilizes tiktoken3\numfootnote{\href{https://github.com/openai/tiktoken/tree/main}{https://github.com/openai/tiktoken/tree/main}}, which provides fast processing and memory efficiency, to tokenize the payload into sub-word units. The tokens generated from Llama3-8B~\cite{dubey2024Llama} are mapped to their corresponding IDs using a vocabulary, similar to Llama2-7B~\cite{touvron2023Llama2}. The padding tokens are then added, and the final sequence is used as input for the CMAE model. The embedding layer of the Llama3-8B model~\cite{dubey2024Llama} is also adopted directly.
    
        \item As depicted in \autoref{fig:subword_tokenizer}, applying tiktoken3 to the UTF-8 converted data results in a list of sub-words such as [b"00", b"c", b"1", …, b"0"] in step 3 (right). The corresponding token IDs retrieved from the vocabulary are [410, 66, 16, …, 15] in Step 4. Finally, this sequence is structured as the model input in Step 5.
    \end{itemize}
\end{itemize}

\subsubsection{\textbf{Strategy 2: Hex Tokenization and Embedding}}\label{sec:hex_token}

\begin{figure}[!htbp]
  \centering
  \includegraphics[height=7cm, keepaspectratio]{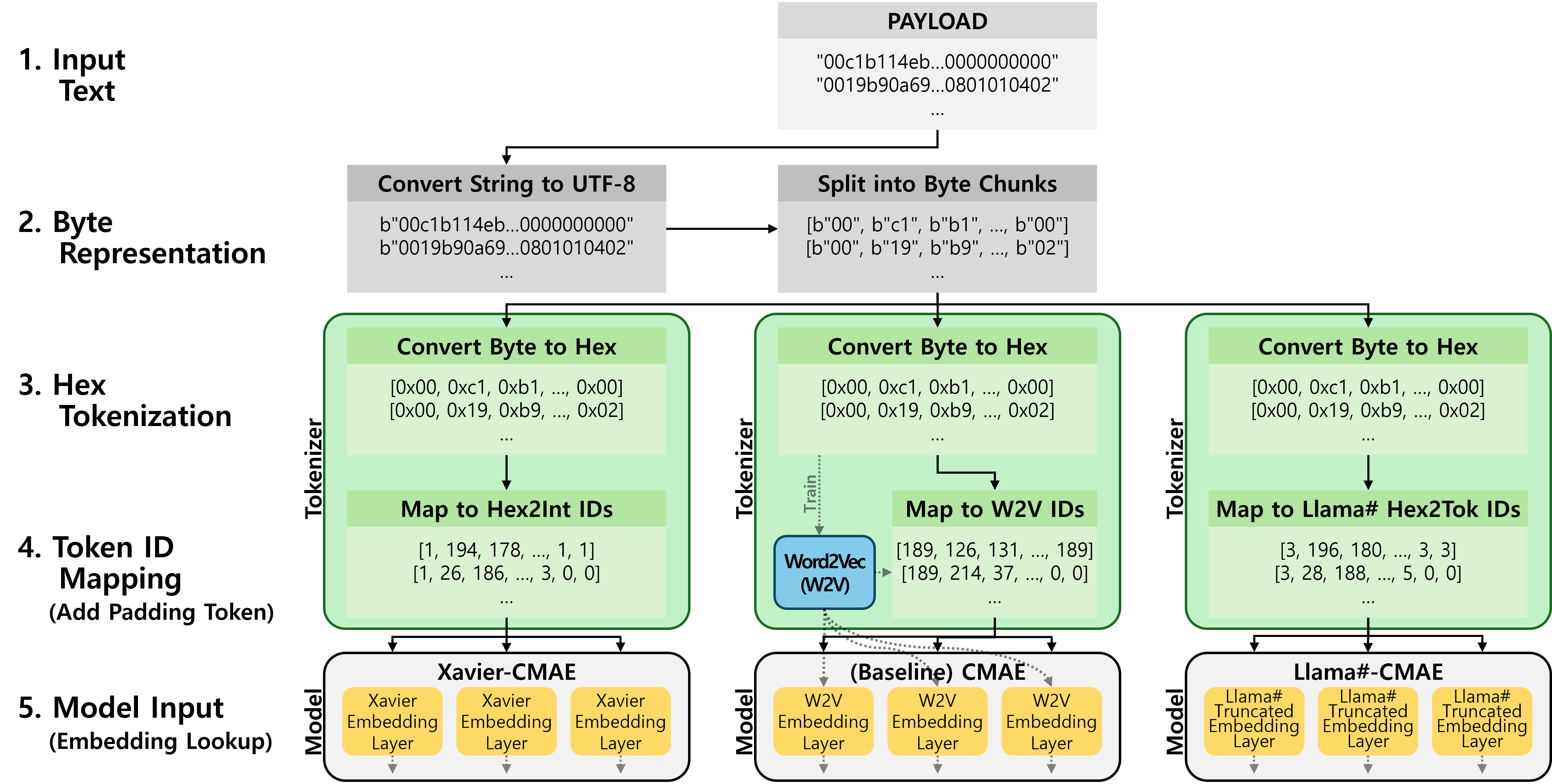}
  \caption{Hex Tokenization}
  \Description{Hex Tokenization}
  \label{fig:hex_tokenizer}
\end{figure}

\leavevmode\newline
The second approach reflects the characteristics of domain-specific data used in the CMAE model~\cite{kim2023ensemble}. As illustrated in steps 1 and 2 of \autoref{fig:hex_tokenizer}, this method converts the input string-type payload data into UTF-8 format and subsequently tokenizes it into fixed 1-byte chunks. Given that UTF-8 is designed to represent all Unicode characters (U+0000 - U+10FFFF), there are no cases where valid Unicode strings cannot be converted. Furthermore, since each byte chunk falls within the range of 0x00 to 0xFF, the total number of tokens is limited to 256 (0x00 - 0xFF), plus one additional padding token, making a total of 257 tokens. This allows for the definition of a scale-free vocabulary and embedding space.

The tokenization process in \autoref{fig:hex_tokenizer}, steps 1 and 2, follows a similar pattern to \autoref{fig:subword_tokenizer}, with an additional step of splitting the input data into fixed-length, two-character segments. For instance, if the payload data is converted to UTF-8 format as b"00c1b114eb…0000000000", it is divided into a list of byte chunks such as [b"00", b"c1", b"b1", …, b"00"]. The tokenizer then processes these byte chunks, and regardless of the model used, they are mapped into hexadecimal values such as [0x00, 0xc1, 0xb1, …, 0x00]. However, the method of mapping word IDs in step 4 and the weight initialization of the embedding layer in Step 5 vary depending on the approach.

\begin{itemize}
    \item \textbf{(Baseline) CMAE Model Based on Word2Vec Tokenizer}
    \begin{itemize}
        \item The approach presented in the middle of \autoref{fig:hex_tokenizer} follows the method used in the conventional CMAE model~\cite{kim2023ensemble}. This method begins with an unsupervised learning phase where a Word2Vec (W2V) model~\cite{mikolov2013efficient} is pre-trained before the main model training. The W2V model~\cite{mikolov2013efficient} generates an embedding space of the same size as the main model's embedding layer, positioning semantically similar words close to each other. The generated vector matrix is then assigned as the weight of the main model’s embedding layer, a common technique in deep learning-based Natural Language Processing (NLP) to optimize initial weights, prevent local optima, and accelerate convergence to the global optimum~\cite{collobert2011natural, erhan2010does}. However, training a W2V model~\cite{mikolov2013efficient} can be time-consuming, especially for large datasets. The vocabulary generated from the W2V model~\cite{mikolov2013efficient} is then used as the tokenizer.

        \item As illustrated in Step 4 of \autoref{fig:subword_tokenizer}, the process first involves training the W2V model~\cite{mikolov2013efficient} to create an embedding matrix and vocabulary. Then, as with other tokenizers, a byte chunk list such as [0x00, 0xc1, 0xb1, …, 0x00] is mapped to token IDs defined in the vocabulary, such as [189, 126, 131, …, 189], before being passed to the model in Step 5.
    \end{itemize}

    \item \textbf{Xavier-CMAE Model Based on Hex2Int Tokenizer}
    \begin{itemize}
        \item The approach shown on the left side of \autoref{fig:hex_tokenizer} is designed to accelerate computation and simplify processing. This method eliminates the W2V model~\cite{mikolov2013efficient} and instead directly constructs a vocabulary. The padding token ID is set to 0, while each input hex value is converted to an integer ID by adding 1 to its decimal equivalent. For example, 0xc1 in hexadecimal corresponds to 193 in decimal, which is mapped to an ID of 194.
        
        \item As shown in Step 4 of \autoref{fig:subword_tokenizer}, when an input of [0x00, 0xc1, 0xb1, …, 0x00] is given, it is converted to [1, 194, 178, …, 1] before being passed to the model in step 5. However, since this method lacks pre-trained embedding vectors, Xavier uniform initialization~\cite{glorot2010understanding} is employed to maintain a consistent output distribution and mitigate vanishing~\cite{rumelhart1986learning} and exploding gradient problems~\cite{lecun1998gradient}.
    \end{itemize}

    \item \textbf{Llama2-CMAE and Llama3-CMAE Model Based on Hex2Tok Tokenizer}
    \begin{itemize}
        \item The method illustrated on the right side of \autoref{fig:hex_tokenizer} leverages pre-trained embedding vectors from LLM models~\cite{touvron2023Llama2, dubey2024Llama} while maintaining the original CMAE model~\cite{kim2023ensemble} architecture. Instead of searching for sub-words, this method selects only the 256 token IDs corresponding to hex values from the vocabulary of pre-trained LLM models~\cite{touvron2023Llama2, dubey2024Llama}. Because this approach does not require maintaining a complete vocabulary, it operates more efficiently.

        \item When using the Llama2-7B model~\cite{touvron2023Llama2}, the 256 tokens corresponding to the range 0x00 to 0xFF and their associated embedding vectors are extracted and integrated into the CMAE model~\cite{kim2023ensemble}. Similarly, when using the Llama3-8B model~\cite{dubey2024Llama}, the same range of tokens and embeddings are utilized.

        \item In Step 4 of \autoref{fig:hex_tokenizer}, a byte chunk list such as [0x00, 0xc1, 0xb1, …, 0x00] is converted into the corresponding token IDs from the defined LLM vocabulary, such as [3, 196, 180, …, 3]. These values are then passed to the model in Step 5 for final processing.
    \end{itemize}
\end{itemize}

\section{Experiment}\label{sec:experiment}

\subsection{Experimental Setup}\label{sec:exp_setup}

\subsubsection{\textbf{Hardware Specifications}}\label{sec:hardware}

\leavevmode\newline
Experiments were conducted on an NVIDIA DGX-1 Server, which is optimized for high-performance deep-learning workloads. The 8 Tesla V100 GPUs enabled efficient parallel processing of large-scale models, while the high-capacity RAM and SSD storage facilitated rapid data loading and training.

\begin{itemize}
    \item \textbf{CPU:} 2x Intel Xeon E5-2698 v4 (2.2 GHz, 20 Cores each)
    \item \textbf{GPU:} 8x NVIDIA Tesla V100 (32 GB VRAM each)
    \item \textbf{RAM:} 512 GB (2,133 MHz DDR4 RDIMM)
    \item \textbf{Storage:} 4x 1.92 TB SSD (RAID 0))
\end{itemize}

\subsubsection{\textbf{Software Environment}}\label{sec:software}

\leavevmode\newline
To ensure environmental consistency and isolation, experiments were conducted within Docker containers. Each container provided a self-contained execution environment, preventing dependency conflicts and allowing for precise control over CUDA, cuDNN, and deep learning frameworks. TensorFlow was used for training the models, while PyTorch was employed for extracting Llama embeddings.

\begin{itemize}
    \item \textbf{Host OS:} Ubuntu 18.04.5 LTS, Docker 24.0.2
    \item \textbf{Container:} Ubuntu 22.04.2 LTS, CUDA 11.8, cuDNN 8.6.0.163-1
    \item \textbf{Frameworks:} Python 3.11.0rc1, TensorFlow 2.13.0 (for training), PyTorch 2.0.1 (for extracting Llama embedding)
\end{itemize}

\subsubsection{\textbf{Dataset Description}}\label{sec:dataset}

\begin{table}[!htbp]
\caption{Class Distribution on the CIC-IDS2017 Dataset}
\label{tab:dataset}
\begin{tabular}{c|cccc}
    \toprule[1.5pt]
    \textbf{Class} & \textbf{Train} & \textbf{validation} & \textbf{Test} & \textbf{Total} \\
    \hline \hline
    Benign & 338,119 & 84,508 & 105,638 & 528,265 \\
    DoS & 160,619 & 40,192 & 49,895 & 250,706 \\
    DDoS & 79,236 & 19,906 & 24,969 & 124,111 \\
    Port Scan & 68,961 & 17,137 & 21,680 & 107,778 \\
    Brute Force & 8,484 & 2,111 & 2,636 & 13,231 \\
    Bot & 1,214 & 315 & 376 & 1,905 \\
    Web & 1,059 & 254 & 335 & 1,648 \\
    \hline \hline
    Total & 657,692 & 164,423 & 205,529 & 1,027,644 \\
    \bottomrule[1.5pt]
\end{tabular}
\end{table}

\leavevmode\newline
CIC-IDS2017~\cite{sharafaldin2018toward} dataset was used for performance and cost evaluation. CIC-IDS2017 is a public network dataset that contains both benign and various attack traffic. The Canadian Institute for Cybersecurity constructed this dataset to capture up-to-date cyber threats and provide a reliable benchmark for intrusion detection research. The dataset was collected over five days (July 3–7, 2017) by simulating real-world attack scenarios and recording network traffic in PCAP format.

For this study, a pre-processed subset of the CIC-IDS2017 dataset was used, where missing values were removed~\cite{alrowaily2019effectiveness}. The dataset consists of 1,027,644 network packets, categorized into seven classes: Benign, DoS, DDoS, Port Scan, Brute Force, Bot, and Web attacks. 

To ensure fair model evaluation, the dataset was split into train, validation, and test sets, preserving the class distribution. First, the dataset was divided into train and test sets at an 8:2 ratio. Then, the training set was further split into train and validation sets at an 8:2 ratio. Thus, the final dataset composition is approximately 64\% train, 16\% validation, and 20\% test. \autoref{tab:dataset} summarizes the number of samples in each set.

\subsection{Model Configuration and Hyperparameters}

\subsubsection{\textbf{Hyperparameters}}

\leavevmode\newline
In this section, we summarize the common settings used for training the models, which include key hyperparameters such as the optimizer, scheduler, loss function, and training configurations like the number of epochs and batch size.

\begin{itemize}
    \item \textbf{Optimizer:} AdaBelief~\cite{zhuang2020adabelief} optimizer was used, with the following hyperparameters applied: learning rate \( 5 \times 10^{-4} \), epsilon \( 1 \times 10^{-16} \), and weight decay \( 1 \times 10^{-4} \).
    \item \textbf{Scheduler:} ReduceLROnPlateau scheduler was used, with the following hyperparameters: factor \( 3 \times 10^{-1} \), patience 2, and minimum learning rate \( 1 \times 10^{-5} \).
    \item \textbf{Early Stopping:} Early stopping with a patience of 5 was applied to halt training if validation performance did not improve.
    \item \textbf{Loss Function:} Categorical CrossEntropy was used.
    \item \textbf{Epochs:} The models were trained for 50 epochs.
    \item \textbf{Batch Size:} The training batch size was fixed at 64, while inference used batch sizes ranging from 8 to 2,048, depending on available memory.
\end{itemize}

The following list provides details on the model architecture, which includes configurations for the embedding layer, convolutional attention blocks, activation functions, and dropout rates. Each model uses different architectural settings to capture various patterns in the data.

\begin{itemize}
    \item \textbf{Word2Vec Embedding:} When Word2Vec is used, the embedding size is 64, with a window size of 5, a minimum count of 5, and the skip-gram model applied.
    \item \textbf{Embedding Layer:} The embedding size varies across models, including 1 padding token: Xavier-CMAE and (Baseline) CMAE~\cite{kim2023ensemble} have an embedding size of \( (257, 64) \), Llama2-CMAE has an embedding size of \( (32,001, 4,096) \), and Llama3-CMAE has an embedding size of \( (128,258, 4,096) \).
    \item \textbf{Convolutional Attention Block:} The Conv1D layers in the Convolutional Attention Block have a kernel size of 3, with filter sizes of 128 and 64 for the two layers, respectively.
    \item \textbf{Attention Block:} The Multi-Head Attention layer in the Attention Block uses \( d_{\text{model}} \) of 64 and \( h \) of 2 attention heads.
    \item \textbf{Multi-Kernel Block:} The Conv1D layers in the Multi-Kernel Block use kernel sizes of 3, 4, and 5 in parallel, with a filter size of 128 for each kernel.
    \item \textbf{Activation Function:} GELU~\cite{hendrycks2016gaussian} was used as the activation function for hidden layers, and Softmax was applied to the output layer.
    \item \textbf{Dropout:} A dropout rate of 0.25 was applied across all models.
\end{itemize}

\begin{table}[!htbp]
    \caption{Number of Parameters by Model and Tokenizer}
    \label{tab:param}
    \begin{tabular}{cc|ccc}
    \toprule[1.5pt]
        \textbf{Model}               & \textbf{Tokenizer} & \textbf{Trainable} & \textbf{Frozen} & \textbf{Total} \\
        \hline \hline
        Xavier-CMAE                  & Hex2Int & 410,823 & 0 & 410,823 \\
        \hline
        (Baseline) CMAE~\cite{kim2023ensemble} & Word2Vec & 410,823 & 0 & 410,823 \\
        \hline
        \multirow{3}{*}{Llama2-CMAE} & BPE+SP\scalebox{1}{*} & 9,651,207 & 393,228,288 & 402,879,495 \\
                                     & Hex2Tok\scalebox{1}{*} & 9,651,207 & 3,158,016 & 12,809,223 \\
                                     & Hex2Tok & 12,809,223 & 0 & 12,809,223 \\
        \hline
        \multirow{3}{*}{Llama3-CMAE} & tiktoken3\scalebox{1}{*} & 9,651,207 & 1,576,022,016 & 1,585,673,223 \\
                                     & Hex2Tok\scalebox{1}{*} & 9,651,207 & 3,158,016 & 12,809,223 \\
                                     & Hex2Tok & 12,809,223 & 0 & 12,809,223 \\
    \bottomrule[1.5pt]
    \end{tabular}
\end{table}

The detailed number of parameters for each model is presented in \autoref{tab:param}. The asterisk (*) next to the tokenizer indicates that the embedding layer is frozen, and if no asterisk is present, the embedding layer is trainable. Since Xavier-CMAE and (Baseline) CMAE~\cite{kim2023ensemble} only differ in the initialization of the embedding layer but share the same configuration overall, they have identical tokenizers and parameter counts. 

Next, for the Llama-CMAE model, the BPE+SP or tiktoken3 tokenizer is frozen, and therefore the original embedding layer is directly concatenated. As a result, it can be observed that it has a large frozen layer. In cases where the Hex2Tok tokenizer is used, there are variations between frozen and unfrozen embedding layers. Due to truncation, the parameter count is significantly lower—approximately 1/20th to 1/100th or less—compared to when using the BPE+SP or tiktoken3 tokenizer. Moreover, despite having the same total number of parameters, the frozen and unfrozen versions differ in the number of trainable parameters.

\subsection{Evaluation Metrics}\label{sec:eval_metric}

The metrics used in this study are as follows. First, the Benign class refers to the traffic that is identified as normal traffic by the IDS. The remaining classes (DoS, DDoS, Port Scan, Brute Force, Bot, and Web) are the classes classified based on the type of attack method. Collectively, these classes are referred to as the Attack classes. The actual label and the predicted label are depicted as $y_i$ and $\hat{y}_i$, respectively.

\begin{itemize}
    \item \textbf{True Positive (TP):} The actual label is true and the predicted label is also true.
        \begin{equation}
        TP = \sum_{i} \left(y_i = true, \hat{y}_i = true \right)
        \end{equation}
    \item \textbf{False Positive (FP):} The actual label is false and the predicted label is true.
        \begin{equation}
        FP = \sum_{i} \left(y_i = false, \hat{y}_i = true \right)
        \end{equation}
    \item \textbf{False Negative (FN):} The actual label is true and the predicted label is false.
        \begin{equation}
        FN = \sum_{i} \left(y_i = true, \hat{y}_i = false \right)
        \end{equation}
    \item \textbf{True Negative (TN):} The actual label is false and the predicted label is also false.
        \begin{equation}
        TN = \sum_{i} \left(y_i = false, \hat{y}_i = false \right)
        \end{equation}

    \item \textbf{Wrongly Detected:} The actual label is the Benign class and the predicted label is one of the Attack classes as used in \autoref{tab:class_distribution}.
        \begin{equation}
        Wrongly \; Detected \; = FP_{Benign} = \sum_{i} \left(y_i = Benign, \hat{y}_i \in Attack \right)
        \end{equation}
    \item \textbf{Missed Attacks:} The actual label is one of the Attack classes and the Predicted label is the Benign class as used in \autoref{tab:class_distribution}.
        \begin{equation}
        Missed \; Attacks \; = FN_{Benign} = \sum_{i} \left(y_i \in Attack, \hat{y}_i = Benign \right)
        \end{equation}

    \item \textbf{Accuracy:} Macro Accuracy as used in \autoref{tab:macro_performance}. (where \( N \) is the number of classes)
        \begin{equation}
        \text{Accuracy} = \frac{1}{N} \sum_{i=1}^{N} \left( \frac{TP_{i} + TN_{i}}{TP_{i} + FP_{i} + FN_{i} + TN_{i}} \right) \times 100 \; (\%)
        \end{equation}
    \item \textbf{Precision:} Macro Precision as used in \autoref{tab:macro_performance}. (where \( N \) is the number of classes)
        \begin{equation}
        \text{Precision} = \frac{1}{N} \sum_{i=1}^{N} \left( \frac{TP_{i}}{TP_{i} + FP_{i}} \right) \times 100 \; (\%)
        \end{equation}
    \item \textbf{Recall:} Macro Recall as used in \autoref{tab:macro_performance}. (where \( N \) is the number of classes)
        \begin{equation}
        \text{Recall} = \frac{1}{N} \sum_{i=1}^{N} \left( \frac{TP_{i}}{TP_{i} + FN_{i}} \right) \times 100 \; (\%)
        \end{equation}
    \item \textbf{F1 Score:} Macro F1 Score as used in \autoref{tab:macro_performance}. (where \( N \) is the number of classes)
        \begin{equation}
        \text{F1 Score} = \frac{1}{N} \sum_{i=1}^{N} \left( 2 \times \frac{Precision_{i} \times Recall_{i}}{Precision_{i} + Recall_{i}} \right) \times 100 \; (\%)
        \end{equation}
    \item \textbf{False Positive Rate (FP Rate):} Macro False Positive Rate as used in \autoref{tab:macro_performance}. (where \( N \) is the number of classes)
        \begin{equation}
        \text{FP Rate} = \frac{1}{N} \sum_{i=1}^{N} \left( \frac{FP_{i}}{FP_{i} + TN_{i}} \right) \times 100 \; (\%)
        \end{equation}
\end{itemize}

\subsection{Experimental Results and Analysis: Detailed Performance}

\begin{table}[!htbp]
    \centering
    \caption{Correct Predictions and Detection Errors on the CIC-IDS2017 Dataset}
    
    \begingroup
    \setlength{\tabcolsep}{3pt} 
    \renewcommand{\arraystretch}{1.1} 
    \resizebox{\textwidth}{!}{ 

        \begin{tabular}{ccc|ccccccc|cc}
            \toprule[1.5pt]
            \textbf{Length} & \textbf{Model} & \textbf{Tokenizer} & 
                \textbf{Benign} & \textbf{Dos} & \textbf{DDoS} & 
                \textbf{\begin{tabular}[c]{@{}c@{}}Port\\[-2pt] Scan\end{tabular}} &
                \textbf{\begin{tabular}[c]{@{}c@{}}Brute\\[-2pt] Force\end{tabular}} &
                \textbf{Bot} & \textbf{Web} & 
                \textbf{\begin{tabular}[c]{@{}c@{}}Wrongly\\[-2pt] Detected\end{tabular}} & 
                \textbf{\begin{tabular}[c]{@{}c@{}}Missed\\[-2pt] Attacks\end{tabular}} \\ 
            \hline \hline
            \multirow{8}{*}{1,500} & Xavier-CMAE & Hex2Int & 105,575 & 49,622 & 23,232 & 20,783 & 2,085 & 329 & \underline{323} & 63 & 1,841 \\ \cline{2-12}
                                   & (Baseline) CMAE~\cite{kim2023ensemble} & Word2Vec & 105,583 & 49,551 & 23,208 & 20,797 & 2,067 & 341 & 316 & 55 & 1,845 \\ \cline{2-12}
                                   & \multirow{3}{*}{Llama2-CMAE} & BPE+SP\scalebox{1}{*} & 105,427 & 49,215 & 23,020 & 20,850 & 1,874 & 317 & 224 & 211 & 1,898 \\
                                   & & Hex2Tok\scalebox{1}{*} & 105,530 & 49,318 & 22,983 & 20,512 & 1,938 & 320 & 241 & 108 & 1,963 \\
                                   & & Hex2Tok & 105,561 & 49,577 & 22,791 & 20,841 & 2,077 & 336 & 306 & 77 & 1,849 \\ \cline{2-12}
                                   & \multirow{3}{*}{Llama3-CMAE} & tiktoken3\scalebox{1}{*} & 105,491 & 48,940 & 24,205 & 20,938 & 1,906 & 311 & 75 & 147 & 1,756 \\
                                   & & Hex2Tok\scalebox{1}{*} & 105,569 & 49,585 & 23,175 & 20,781 & 2,062 & 335 & 309 & 69 & 1,880 \\
                                   & & Hex2Tok & 105,587 & 49,616 & 23,218 & 20,762 & 2,068 & 334 & 306 & 51 & 1,878 \\
           \hline
           \multirow{8}{*}{3,000} & Xavier-CMAE & Hex2Int & 105,603 & 49,347 & 24,833 & 21,514 & 2,536 & \underline{360} & 311 & 35 & 268 \\ [1.5pt] \cline{2-12}
                                   & (Baseline) CMAE~\cite{kim2023ensemble} & Word2Vec & 105,548 & 49,417 & 24,862 & 21,554 & 2,579 & 357 & \underline{320} & 90 & 181 \\ \cline{2-12}
                                   & \multirow{3}{*}{Llama2-CMAE} & BPE+SP\scalebox{1}{*} & 104,088 & 49,166 & 24,772 & 21,428 & 2,488 & 339 & 245 & 1,550 & 220 \\
                                   & & Hex2Tok\scalebox{1}{*} & 105,580 & 49,327 & 24,740 & 21,347 & 2,516 & 355 & 306 & 58 & 238 \\
                                   & & Hex2Tok & 105,587 & 49,383 & 24,855 & 21,550 & 2,572 & 355 & 318 & 51 & 225 \\ \cline{2-12}
                                   & \multirow{3}{*}{Llama3-CMAE} & tiktoken3\scalebox{1}{*} & 105,592 & 49,667 & 24,904 & \underline{21,624} & 2,600 & \textbf{\underline{363}} & 313 & 46 & \textbf{\underline{25}} \\
                                   & & Hex2Tok\scalebox{1}{*} & 105,586 & 49,388 & 24,858 & 21,545 & 2,567 & 361 & 315 & 52 & 221 \\
                                   & & Hex2Tok & 105,566 & 49,385 & 24,862 & 21,534 & 2,569 & 338 & 311 & 72 & 220 \\ 
            \hline
            \multirow{6}{*}{Max} & Xavier-CMAE & Hex2Int & \underline{105,618} & \textbf{\underline{49,851}} & \textbf{\underline{24,934}} & \underline{21,620} & \underline{2,618} & \underline{361} & \textbf{\underline{324}} & \underline{20} & \underline{30} \\ [1.5pt] \cline{2-12}
                                 & (Baseline) CMAE~\cite{kim2023ensemble} & Word2Vec & 105,595 & 49,818 & 24,915 & \textbf{\underline{21,635}} & \textbf{\underline{2,622}} & 359 & \underline{321} & 43 & 32 \\ [1.5pt] \cline{2-12}
                                 & \multirow{2}{*}{Llama2-CMAE} & BPE+SP\scalebox{1}{*} & 105,537 & 49,760 & 24,881 & 21,541 & 2,570 & 344 & 284 & 101 & 41 \\
                                 & & Hex2Tok\scalebox{1}{*} & 105,561 & 49,752 & 24,729 & 21,484 & 2,535 & 357 & 308 & 77 & 51 \\ [1.5pt] \cline{2-12}
                                 & \multirow{2}{*}{Llama3-CMAE} & tiktoken3\scalebox{1}{*} & 105,576 & 49,795 & 24,911 & 21,586 & 2,596 & 357 & 286 & 62 & 29 \\
                                 & & Hex2Tok\scalebox{1}{*} & \textbf{\underline{105,622}} & \underline{49,850} & \underline{24,922} & \underline{21,622} & \underline{2,617} & \underline{361} & 316 & \textbf{\underline{16}} & 32
            \\ [1.5pt]
            \hline \hline
            \multicolumn{3}{c|}{Number of Samples} & 105,638 & 49,895 & 24,969 & 21,680 & 2,636 & 376 & 335 & 105,638 & 99,891 \\
            \bottomrule[1.5pt]
        \end{tabular}
    }
    \endgroup
    \label{tab:class_distribution}
\end{table}

\autoref{tab:class_distribution} presents the number of correct predictions and detection errors in the CIC-IDS2017 test dataset based on payload input length, model type, and tokenizer type. Also \autoref{tab:class_distribution} and \autoref{tab:macro_performance} follow the default settings:
\begin{itemize}
    \item During training, eight NVIDIA Tesla V100 GPUs (32GB each) were used with a batch size of 64. Lighter models, such as Xavier-CMAE (Hex2Int tokenizer, payload truncated to 1,500) and CMAE~\cite{kim2023ensemble} (Word2Vec tokenizer, payload truncated to 1,500), required only 2GB of VRAM per GPU, allowing can be use larger batches. In contrast, LLM-CMAE models (full-length payloads) consumed significantly more memory, reaching 32GB per GPU.

    \item For inference, a single GPU with 32GB VRAM was used. Within this constraint, lighter models supported larger batch sizes, reaching 2,048 before hitting memory limits. Meanwhile, LLM-CMAE models, which process full-length payloads, were limited to a batch size of 8 due to their higher memory footprint.
\end{itemize}

\subsubsection{\textbf{Comparison of Payload Length on Detection Performance}}

\leavevmode\newline
First, when analyzing performance based on payload length, it was observed that using longer payloads yielded higher accuracy than truncating the input to 1,500 bytes. Although most attack packets are concentrated within the first 1,500 bytes, certain attack packets extend beyond this range, potentially evading detection. Notably, increasing the payload length from 1,500 to 3,000 bytes reduced the number of missed attacks to approximately one-ninth of the original value.

However, increasing payload length not only degrades computational efficiency but also leads to a sharp rise in memory consumption. Using the full payload length without truncation may cause Out-of-Memory (OOM) issues. For instance, in the case of the Llama-CMAE model, when the payload length is set to the maximum value, the embedding layer cannot be trained due to memory constraints.

\subsubsection{\textbf{Comparison of Tokenizer on Model Detection Performance}}

\leavevmode\newline
When analyzing performance based on tokenizer type, it was found that Hex2Int outperformed Word2Vec as payload length increased. While Word2Vec provides useful initial embedding vectors for short payloads, longer payloads benefit more from Xavier initialization, which distributes embeddings with appropriate variance, allowing for better convergence to a global optimum compared to Word2Vec.

Furthermore, although the Hex2Tok method generally outperformed sub-word tokenization (BPE+SP or tiktoken3), there were exceptions where sub-word approaches were more effective. Specifically, when truncating the payload to 3,000 bytes, the Llama3-CMAE model using the tiktoken3 tokenizer achieved the highest accuracy for detecting Bot traffic and reducing missed attacks.

\subsubsection{\textbf{Comparison of Model Detection Performance}}

\leavevmode\newline
When comparing model performance, the Llama3-CMAE model consistently outperformed the Llama2-CMAE model, indicating that a superior pre-trained backbone contributes to improved embedding representations and overall detection accuracy.

Despite the high overall predictive performance, the dataset exhibits an imbalanced class distribution, leading to variations in detection performance across different attack types. Since even a single missed attack can result in significant security threats, it is crucial to evaluate model performance not only at an aggregate level but also for each specific attack category. Depending on the operational security environment and detection latency requirements, certain applications may require low-latency, high-precision models specialized for specific attack types. In such cases, downstream classification tasks or separate sub-models may be necessary to enhance service robustness. The detection performance by attack type is summarized as follows:

\begin{itemize}
    \item \textbf{Benign Traffic:} The Llama3-CMAE model with a frozen Hex2Tok tokenizer achieved the highest classification accuracy when using the maximum payload length.
    \item \textbf{DoS/DDoS and Web Attacks:} The Xavier-CMAE model with a Hex2Int tokenizer demonstrated the highest detection accuracy when using the maximum payload length.
    \item \textbf{Port Scanning and Brute Force Attacks:} The baseline CMAE model~\cite{kim2023ensemble} with a Word2Vec tokenizer achieved the best performance when using the maximum payload length.
    \item \textbf{Bot Attacks:} The Llama3-CMAE model with a frozen tiktoken3 tokenizer achieved the highest detection performance when using a truncated payload length of 3,000 bytes.
\end{itemize}

\subsubsection{\textbf{Comparison of Wrongly Detected and Missed Attacks}}

\leavevmode\newline
From the perspective of Wrongly Detected cases, the Llama2-CMAE model with a 3,000-byte payload length exhibited a tendency to misclassify benign traffic as attacks, indicating that the model likely failed to escape a local optimum. In contrast, the Llama3-CMAE model using the maximum payload length recorded only 16 wrongly detected instances, demonstrating the most stable detection performance.

Regarding Missed Attacks, the Llama3-CMAE model using the tiktoken3 tokenizer with a 3,000-byte payload length exhibited the lowest number of missed attacks (25 instances), confirming its effectiveness in detecting malicious packets.

\subsection{Experimental Results and Analysis: Macro-Level Performance}

\begin{table}[!htbp]
    \centering
    \caption{Macro Performance Comparison on the CIC-IDS2017 Dataset}
    \begingroup
    \setlength{\tabcolsep}{3pt} 
    
    \renewcommand{\arraystretch}{1.1} 
    \resizebox{\textwidth}{!}{ 
        \begin{tabular}{ccc|ccccc|cc}
            \toprule[1.5pt]
            \textbf{Length} & \textbf{Model} & \textbf{Tokenizer} &
            \textbf{\begin{tabular}[c]{@{}c@{}}Accuracy\\[-2pt] (\%)\end{tabular}} & 
            \textbf{\begin{tabular}[c]{@{}c@{}}Precision\\[-2pt] (\%)\end{tabular}} & 
            \textbf{\begin{tabular}[c]{@{}c@{}}Recall\\[-2pt] (\%)\end{tabular}} & 
            \textbf{\begin{tabular}[c]{@{}c@{}}F1 Score\\[-2pt] (\%)\end{tabular}} & 
            \textbf{\begin{tabular}[c]{@{}c@{}}FP Rate\\[-2pt] (\%)\end{tabular}} & 
            \textbf{\begin{tabular}[c]{@{}c@{}}Train Time\\[-2pt] (Hours)\end{tabular}} & 
            \textbf{\begin{tabular}[c]{@{}c@{}}Predict Time\\[-2pt] (Predict/s)\end{tabular}} \\
            \hline \hline
            \multirow{8}{*}{1,500} & Xavier-CMAE & Hex2Int & \textbf{\underline{99.5023}} & 97.9002 & 93.0449 & 95.2268 & 0.4226 & \textbf{\underline{2.1}} & \textbf{\underline{5,559.3}} \\ \cline{2-10}
             & (Baseline) CMAE~\cite{kim2023ensemble} & Word2Vec & 99.4904 & 98.3899 & \textbf{\underline{93.0810}} & \textbf{\underline{95.5002}} & 0.4295 & 3.8 & 5,021.5 \\ \cline{2-10}
             & \multirow{3}{*}{Llama2-CMAE} & BPE+SP\scalebox{1}{*} & 99.3603 & 94.9995 & 87.0100 & 90.5040 & 0.5054 & 9.3 & 217.3 \\
             & & Hex2Tok\scalebox{1}{*}  & 99.3484 & 91.9016 & 87.9953 & 89.5408 & 0.5164 & 10.3 & 428.9 \\
             & & Hex2Tok & 99.4384 & 97.0827 & 92.3137 & 94.4883 & 0.4644 & 26.4 & 444.4 \\ \cline{2-10}
             & \multirow{3}{*}{Llama3-CMAE} & tiktoken3\scalebox{1}{*} & 99.4908 & 96.5243 & 81.2674 & 85.6263 & \textbf{\underline{0.4082}} & 17.6 & 212.7 \\
             & & Hex2Tok\scalebox{1}{*} & 99.4838 & 97.4607 & 92.5058 & 94.7502 & 0.4353 & 10.4 & 422.3 \\
             & & Hex2Tok & 99.4943 & \textbf{\underline{98.7752}} & 92.3958 & 95.3258 & 0.4302 & 26.5 & 436.2 \\ 
             \hline
            \multirow{8}{*}{3,000} & Xavier-CMAE & Hex2Int & 99.8575 & 97.1723 & 97.4779 & 97.2811 & 0.0979 & 2.8 & 2,795.6 \\ \cline{2-10}
             & (Baseline) CMAE~\cite{kim2023ensemble} & Word2Vec & 99.8760 & 98.5423 & 98.0363 & 98.2754 & 0.0822 & 5.1 & 2,564.9  \\ \cline{2-10}
             & \multirow{3}{*}{Llama2-CMAE} & BPE+SP\scalebox{1}{*} & 99.5825 & 96.1944 & 93.2571 & 94.5575 & 0.2511 & 31.5 & 111.2 \\
              & & Hex2Tok\scalebox{1}{*} & 99.8112 & 96.0786 & 96.7942 & 96.3862 & 0.1221 & 17.0 & 216.4 \\
              & & Hex2Tok & 99.8736 & 98.5152 & 97.8260 & 98.1583 & 0.0862 & 47.4 & 223.1 \\ \cline{2-10}
             & \multirow{3}{*}{Llama3-CMAE} & tiktoken3\scalebox{1}{*} & \underline{99.9352} & 98.4221 & \underline{98.2272} & 98.3213 & \underline{0.0382} & 31.1 & 108.9 \\
              & & Hex2Tok\scalebox{1}{*} & 99.8736 & \underline{98.9354} & 97.8986 & \underline{98.4078} & 0.0859 & 17.1 & 207.2 \\
              & & Hex2Tok & 99.8660 & 98.7612 & 96.8565 & 97.7662 & 0.0905 & 47.4 & 213.4 \\
             \hline
            \multirow{6}{*}{Max} & Xavier-CMAE & Hex2Int & \textbf{\underline{99.9718}} & \textbf{\underline{99.6410}} & \textbf{\underline{98.7886}} & \textbf{\underline{99.2082}} & \textbf{\underline{0.0182}} & \textbf{\underline{9.1}} & \textbf{\underline{590.7}} \\ \cline{2-10}
             & (Baseline) CMAE~\cite{kim2023ensemble} & Word2Vec & 99.9633 & 99.0351 & \underline{98.5928} & 98.8068 & 0.0229 & 11.1 & 552.8 \\ \cline{2-10}
             & \multirow{2}{*}{Llama2-CMAE} & BPE+SP\scalebox{1}{*} & 99.9149 & 98.0914 & 96.0574 & 97.0371 & 0.0512 & 122.1 & 23.5 \\
              & & Hex2Tok\scalebox{1}{*} & 99.8884 & 96.0236 & 97.2615 & 96.5594 & 0.0664 & 64.9 & 47.5 \\ \cline{2-10}
             & \multirow{2}{*}{Llama3-CMAE} & tiktoken3\scalebox{1}{*} & 99.9399 & 98.5168 & 96.7855 & 97.6246 & 0.0362 & 113.5 & 24.8 \\
              & & Hex2Tok\scalebox{1}{*} & \underline{99.9696} & \underline{99.4563} & 98.4367 & \underline{98.9374} & \underline{0.0194} & 65.1 & 46.9 \\
             \bottomrule[1.5pt]
        \end{tabular}
    }
    \endgroup
    \label{tab:macro_performance}

\end{table}

\footnotetext[1]{An asterisk (\scalebox{1}{*}) next to a tokenizer name indicates that the embedding layer is frozen. Tokenizers without this mark use a trainable embedding layer.}

\autoref{tab:macro_performance} presents the macro performance results for the CIC-IDS2017 test dataset, based on payload input length, model type, and tokenizer type.

\subsubsection{\textbf{Comparison of Payload Length on Detection Performance}}

\leavevmode\newline
Truncating the payload length to 1,500 significantly improved inference speed, making it well-suited for high-traffic environments. Among the models, Xavier-CMAE achieved the highest accuracy, while CMAE~\cite{kim2023ensemble} demonstrated the best F1 score. Notably, Xavier-CMAE also had the fastest training time (2.1 hours) and inference speed (5,559 packets per second), processing data 26 times faster than Llama3-CMAE, which used the slowest tiktoken3 tokenizer. Furthermore, compared to Llama2-CMAE with BPE+SP, which processed full-length payloads, Xavier-CMAE was 236 times faster.

Increasing the payload length to 3,000 led to Llama3-CMAE achieving the highest accuracy. However, it still underperformed in both accuracy and computational efficiency compared to Xavier-CMAE using the full-length payload. Compared to Xavier-CMAE with a 1,500 payload, Llama3-CMAE showed a 0.35 percentage point accuracy gain but suffered from a 50 \% reduction in inference speed, indicating a clear trade-off. This suggests that Llama3-CMAE may be a viable option when full-length payload processing is infeasible due to resource limitations.

Without truncation, using the full payload length, Xavier-CMAE demonstrated the best overall performance across all macro evaluation metrics. It outperformed all Llama-CMAE variants in both training and inference speed. Compared to its 1,500-payload version, real-time throughput decreased by a factor of nine, but accuracy improved by 0.47 percentage points, reaching 99.97 \%, setting a new state-of-the-art benchmark.

\subsubsection{\textbf{Comparison of Wrongly Detected and Missed Attacks}}

\leavevmode\newline
Analyzing Missed Attacks from \autoref{tab:class_distribution}, the Xavier-CMAE model with full payload length reduced the number of missed attacks by a factor of 60 compared to the 1,500-byte truncated version, demonstrating a significant improvement in attack detection capabilities. Additionally, the number of Wrongly Detected cases was reduced to one-third, indicating enhanced model stability.

A reduction in missed attacks means that network security analysts have a greater opportunity to identify and mitigate emerging threats. Simultaneously, a reduction in false positives (wrongly detected cases) reduces unnecessary workload, allowing security teams to allocate more resources toward detecting new attack patterns and strengthening overall network defense strategies.

\section{Conclusion}\label{sec:conclusion}

\subsection{Summary and Discussion}\label{sec:summary}

In this study, we proposed a CMAE-based model that effectively detects malicious traffic by learning the characteristics of payloads. We applied various tokenizers and training strategies to compare and analyze model performance. Specifically, we introduced the Xavier-CMAE model, which improves upon the conventional Word2Vec-based CMAE~\cite{mikolov2013efficient} by maintaining a simpler architecture while achieving superior detection performance.

Experimental results revealed that the length of the payload significantly impacts detection performance. Using untruncated payloads or increasing the maximum payload length consistently improved accuracy and drastically reduced missed attacks compared to truncating payloads at 1,500 tokens. Additionally, our findings demonstrated that tokenizer selection has a substantial effect on both detection accuracy and false positive rates.

The key findings of our study are as follows:

\begin{itemize}
    \item The proposed Xavier-CMAE model outperformed all other models across macro performance metrics, achieving an accuracy of 99.97\% and an FP rate of 0.0182\%, setting a new state-of-the-art benchmark.

    \item In resource-constrained environments, the Xavier-CMAE model with a 1,500-token payload exhibited the best balance between computational efficiency and accuracy, outperforming other models using the same payload length.

    \item In scenarios where even a single missed attack could lead to severe consequences, a parallel evaluation strategy leveraging multiple models tailored to different attack types could enhance the detection of high-risk packets.
\end{itemize}

\subsection{Limitations}\label{sec:limit}

Despite its promising results, this study has certain limitations:

\begin{itemize}

    \item \textbf{Memory consumption and computational cost:} When using untruncated payloads, models such as Llama-CMAE required substantial VRAM, preventing the application of trainable embedding layers and necessitating the use of frozen embeddings. This constraint may reduce the model’s adaptability and limit its practicality in environments with restricted GPU resources.

    \item \textbf{Generalization to diverse attack types:} While CIC-IDS2017 includes major network attack categories, it does not account for emerging malicious traffic patterns or mutation attacks. Furthermore, the robustness of the model against adversarial attacks remains an open question that requires further investigation.
\end{itemize}

\subsection{Future Work}\label{sec:future_work}

To address these limitations and enhance the practical utility of our model, we suggest the following directions for future research:

\begin{itemize}
    \item \textbf{Extension to real-world and emerging network security threats:} Unlike prior studies that rely on pre-extracted features, our approach necessitates direct packet extraction from raw pcap files. While several publicly available datasets provide raw pcap files, they often lack comprehensive labeling, requiring an additional preprocessing stage. This preprocessing step is both labor-intensive and computationally demanding, posing challenges for large-scale evaluations across multiple datasets within the scope of this study. Furthermore, the acquisition of real-world network traffic data remains a significant challenge due to security and privacy concerns, further limiting the availability of labeled datasets. Future research will focus on extending our methodology to additional datasets such as CSE-CIC-IDS2018, UNSW-NB15, and Bot-IoT, contingent on the development of efficient labeling and preprocessing techniques. This will facilitate a more comprehensive evaluation of the model’s adaptability and robustness across diverse and dynamic network environments.

    \item \textbf{Leveraging state-of-the-art Open LLMs for intrusion detection:} The rapid evolution of Open LLMs (Large Language Models) presents new opportunities for security applications. Comparative studies between Llama2~\cite{touvron2023Llama2}, Llama3~\cite{dubey2024Llama}, and newer models such as Llama3.2 and Llama3.3 could offer valuable insights. Additionally, evaluating alternative open-source LLMs such as Bloom~\cite{le2023bloom}, Falcon~\cite{almazrouei2023falcon}, Mixtral~\cite{jiang2024mixtral}, and MPT~\cite{basyal2023text} could determine whether more advanced language models improve security performance. Finally, domain-specific pretraining and fine-tuning techniques could be explored to further optimize LLM-based intrusion detection systems.

    \item \textbf{Developing memory-efficient model architectures:} Using untruncated payloads increases both computational load and memory consumption. While our study maintained structural consistency with prior work~\cite{kim2023ensemble}, further optimizations—such as consolidating multiple embedding layers into a single one—were not explored. Future research should focus on designing lightweight architectures that preserve accuracy while improving inference speed through techniques such as pruning, quantization, and knowledge distillation.

    \item \textbf{Enhancing robustness against adversarial attacks:} AI-based security systems are increasingly vulnerable to adversarial perturbations and evasion attacks. Future work could explore adversarial training or other robustness-enhancing techniques to mitigate these risks.

\end{itemize}

By addressing these challenges, we aim to develop a more efficient, generalizable, and robust payload-based intrusion detection system for modern network environments.

\begin{acks}
This work was supported in part by the MSIT (Ministry of Science and ICT), Korea under the ITRC (Information Technology Research Center) support program (IITP-2025-RS-2023-00259099, 50\%) supervised by the IITP (Institute for Information \& Communications Technology Planning \& Evaluation, and by the National Research Foundation of Korea(NRF) grant funded by the Korea government(MSIT) (No. RS-2023-00240211, 50\%).
\end{acks}

\bibliographystyle{ACM-Reference-Format}
\bibliography{sections/references.bib}



\appendix
\end{document}